\documentclass[aps,prl,twocolumn,superscriptaddress,groupedaddress,showpacs]{revtex4}
\usepackage{graphicx,epsf}
\usepackage{psfrag}
\begin{document}
\title{Multifractality of self-avoiding walks on percolation clusters}
\author{Viktoria Blavatska}
\email[]{E-mail: viktoria@icmp.lviv.ua; blavatska@itp.uni-leipzig.de}
\affiliation{Institut f\"ur Theoretische Physik and Centre for Theoretical Sciences (NTZ),\\ Universit\"at Leipzig, Postfach 100\,920,
D-04009 Leipzig, Germany}
\affiliation{Institute for Condensed
Matter Physics of the National Academy of Sciences of Ukraine,\\
79011 Lviv, Ukraine}
\author{Wolfhard Janke}
\email[]{E-mail: Wolfhard.Janke@itp.uni-leipzig.de}
\affiliation{Institut f\"ur Theoretische Physik and Centre for Theoretical Sciences (NTZ),\\ Universit\"at Leipzig, Postfach 100\,920,
D-04009 Leipzig, Germany}
 
\begin{abstract}

We consider self-avoiding walks (SAWs) on the backbone of percolation clusters in space dimensions $d=2, 3, 4$.  
Applying numerical simulations, we show that the whole multifractal spectrum of singularities emerges in exploring the peculiarities of the model.  
We obtain estimates for the set of critical exponents, 
that govern scaling laws of higher moments of the distribution of percolation cluster sites visited by SAWs, in a good correspondence with an appropriately summed 
field-theoretical $\varepsilon=6-d$-expansion (H.-K. Janssen and O. Stenull, Phys. Rev. E 75, 020801(R)$\,$(2007)).
   
\end{abstract}
\date{\today}
\maketitle%

When studying physical processes on complicated fractal objects, one often encounters 
the interesting situation of coexistence of a family of singularities, 
each associated with a set of different fractal dimensions \cite{Stanley88}. 
In these problems, the 
conventional scaling approach cannot describe the system. Instead, an infinite set of critical exponents 
is needed to characterize the different moments of the distribution of observables, which scale independently.
These peculiarities are usually referred to as multifractality \cite{Mandelbrot74}.  The multifractal spectrum 
can be used to provide information on the subtle geometrical properties of a fractal object, which cannot be fully described by its fractal dimensionality. Indeed, clusters generated by diffusion-limited aggregation (DLA) \cite{Witten81} and percolation clusters have the same fractal dimensions, but a completely different geometrical structure, which can be clarified, e.g., by studying the multifractality of the voltage distribution in percolation clusters \cite{Arcangelis85} and the growth probability distribution in DLA \cite{Halsey86,Jensen02}.
Multifractal properties arise also in many different contexts, for example in studies of turbulence in chaotic dynamical systems and 
strange attractors \cite{Mandelbrot74,Hentschel83}, human heartbeat dynamics \cite{Ivanov99}, Anderson localization transition \cite{Schreiber91}
etc.

To understand the common roots underlying this phenomenon, it is worthwhile to consider the generic case of self-avoiding walks (SAWs) on fractal clusters. 
It is well established  that configurational properties of SAWs on a regular lattice are governed by scaling laws; 
e.g. for the averaged end-to-end distance $\langle r \rangle$ of a SAW with $N$ steps one finds in 
the asymptotic limit $N\to\infty$:
\begin{equation}
\langle r \rangle \sim N^{\nu_{{\rm SAW}}},
\end{equation}
where $\nu_{{\rm SAW}}$ is an universal exponent depending only on the space dimension $d$.  

The scaling of SAWs changes crucially, when the underlying lattice has a fractal structure.  Indeed, new critical exponents were found for SAWs residing, e.g., on a Sierpinski gasket and Sierpinski carpet \cite{Dhar78}. 
A related problem arises when studying SAWs on disordered lattices with concentration $p$ of structural defects very close to the percolation threshold $p_c$. In this case, an incipient cluster of pure sites can be found in the system. The diameter of a typical  cluster below $p_c$ is characterized by the correlation length $\xi$, which diverges as
$ \xi \sim (p-p_c)^{-\nu_{p}}$ 
with an universal exponent $\nu_{p}$. 
Note  that percolation clusters are fractal objects (see Table~1) and apparently change the universality class of residing SAWs; the scaling (1) holds in this case with an exponent $\nu_{p_{c}} \neq \nu_{{\rm SAW}}$. 
Aiming to study the scaling of SAWs on a percolative lattice, we are interested rather in the backbone of  percolation cluster:
the structure left when all ``dangling ends" are eliminated from the cluster. 
Infinitely long chains can 
only exist on the backbone of the cluster.

\begin{table}[t!]
\small{
\caption{Fractal
dimensions of percolation cluster $d_{p_c}^F$ and backbone of the percolation cluster $d_{p_c}^B$, 
 and correlation length critical exponent $\nu_p$ for different space dimensions $d$.}
\begin{tabular}{rccc}
\hline \hline
$d$ & $d_{p_c}^F$ & $d_{p_c}^B$ & $\nu_p$ \\ 
\hline 
2 &  $91/49$ {\cite{Havlin87}}  & $1.650\pm0.005$ {\cite{Moukarzel98}}  & 4/3 \cite{Nijs79} \\ 
3  &  $2.51\pm0.02 $ {\cite{Grassberger86}} &  $1.86\pm0.01$ {\cite{Moukarzel98}}  & $0.875\pm0.008$ \cite{Strenski91}\\ 
4 &  $3.05\pm0.05$ {\cite{Grassberger86}}  &  $1.95\pm 0.05 $  {\cite{Moukarzel98}}  & $0.69\pm 0.05$ \cite{Moukarzel98}\\ 
 \hline 
\hline
\end{tabular}
\label{dim}
}
\end{table}

Although the behavior of SAWs on percolative lattices served as a subject of numerous numerical \cite{Kremer81, Meir89, Grassberger93,Rintoul94,Ordemannsaw,Blavatska08} and analytical \cite{Kim83,Barat91,Blavatska04,Janssen07} studies since the early 80th, 
not enough attention has been paid to clarifying the multifractality of the problem. 
Following an early idea of Meir and Harris \cite{Meir89},  it was only recently proven in field-theoretical studies \cite{Blavatska04,Janssen07} 
that the exponent $\nu_{p_c}$ alone is not sufficient to completely describe the peculiarities of SAWs on percolation clusters. 
Instead, a whole spectrum $\nu^{(q)}$ of multifractal exponents emerges \cite{Janssen07}:
 \begin{eqnarray}
\nu^{(q)}{=}\frac{1}{2}{+}\left(\!\frac{5}{2}{-}\frac{3}{2^q}\!\right)\frac{\varepsilon}{42}{+}\left(\!\frac{589}{21}{-}\frac{397}{14\cdot2^q}{+}\frac{9}{4^q} \!\right)\left( \frac{\varepsilon}{42}\right)^2, \label{spectr}
\end{eqnarray} 
with $\varepsilon=6-d$. 
Note that putting $q=0$ in (\ref{spectr}), we restore an estimate for the dimension $d_{p_c}^B$ of the underlying backbone of percolation clusters found previously \cite{Janssen00} via $\nu^{(0)}=1/d_{p_c}^B$,
 whereas $q\to \infty$ gives the percolation correlation length exponent $\nu^{(\infty)}=\nu_p$, which restores, e.g., the result of Amit \cite{Amit76}.
$\nu^{(1)}$ gives us the exponent $\nu_{p_c}$, governing the scaling law for the averaged end-to-end distance of SAWs on the backbone of percolation clusters.
This is  a major step forward, but the reliability of a two-loop  
$\varepsilon$-expansion in $\varepsilon=6-d$ remains a priori questionable in physical dimensions $d=2,3$  where $\varepsilon$ is large. To settle this question,
we report in this letter a careful computer simulation study of SAWs on percolation clusters.
We primarily aim to obtain precise numerical estimates for the multifractal exponents $\nu^{(q)}$ in a wide range of $q$ and find the related spectrum of singularities on 
the underlying fractal cluster, thus going into deeper understanding of peculiarities of the model. 

\begin{figure}[t!]
 \begin{center}
\includegraphics[width=4cm]{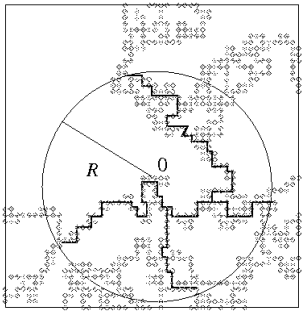}
\end{center}
\caption{\label{multi} Different SAW trajectories with fixed end-to-end distance $R$ on the backbone of a percolation cluster in $d=2$. }
\end{figure}

We consider site percolation on regular lattices of edge lengths up to $L_{{\rm max}}{=}400,200,50$ in dimensions $d{=}2,3,4$, respectively. Each site of 
the lattice is occupied randomly with probability $p_c$ and empty otherwise.
If for a given lattice it is not possible to find a spanning cluster, this disordered lattice is rejected and a new one is constructed. To  extract 
the backbone of a percolation cluster,  we apply the algorithm  proposed in Ref.~\cite{Porto97}. First, the starting point -- ``seed'' -- is chosen at the center of the cluster.  
Then, for all the sites on the borders of the lattice, the shortest paths  to the  ``seed"  are found, forming the so-called elastic backbone \cite{Havlin84}. 
Finally, considering successively each site of the elastic backbone and checking whether a ``loop" (path of sites connected to the elastic backbone in two places at 
least) starts on it, we obtain the geometric backbone of the cluster. We construct $1000$ clusters in each space dimension.  

Starting on the ``seed" of a single cluster, we construct a SAW on it, applying the pruned-enriched chain-growth algorithm \cite{Rosenbluth55}.
 We let a trajectory of SAW grow step by step, until it reaches some prescribed distance (say $R$) from the starting point. Then, the algorithm is stopped,  
and a new SAW grows from the same starting point. In such a way, we are interested in constructing different possible trajectories with fixed end-to-end distance,
 as is shown schematically  in Fig.~\ref{multi}. For each lattice size $L$, we change $R$ up to $\approx L/3$ to avoid finite-size effects, since close to 
lattice borders the structure of the backbone of percolation clusters is distorted and thus can falsify the SAW statistics.

 \begin{figure}[t!]
 \begin{center}
\includegraphics[width=5cm]{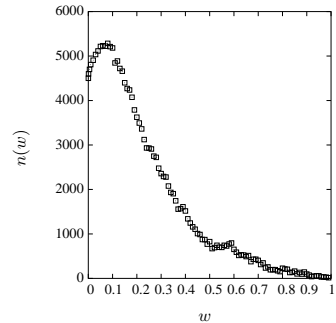}
\end{center}
\caption{\label{dist} Number of sites $n(w)$ of the backbone of a percolation cluster with weights $w$, visited by SAWs with fixed end-to-end distance $R=160$, in $d=2$ dimension.}
\end{figure}

Let us denote by $K(R)$ the total number of constructed SAW trajectories between $0$ and $R$ (we perform $\sim10^6$ SAWs for each value of $R$). 
Then, for each site $i$ of the backbone  we sum up the portion of trajectories, passing through this site. In such a way, we prescribe a weight $w(i)=K(i)/K(R)$ 
to each site $i\in R$, and thus receive some ``population", occupying the underlying fractal cluster. 
The distribution $n(w)$ of weights, prescribed to the sites of the backbone of a percolation cluster, visited by SAWs with fixed end-to-end distance $R=160$ in two 
dimensions, is shown in Fig.~\ref{dist}. 

\begin{figure}[t!]
 \begin{center}
\includegraphics[width=4.3cm,height=3.7cm]{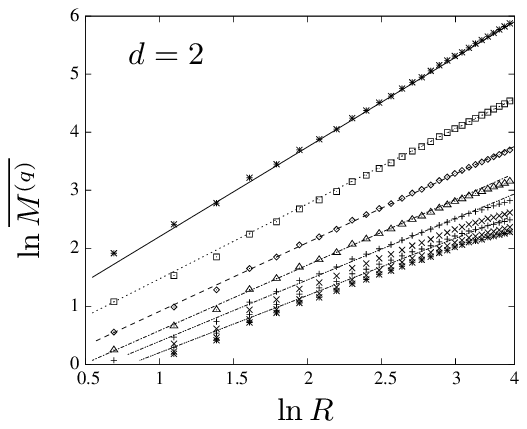}\\
\hspace*{0.1cm}
\includegraphics[width=4.3cm,height=3.7cm]{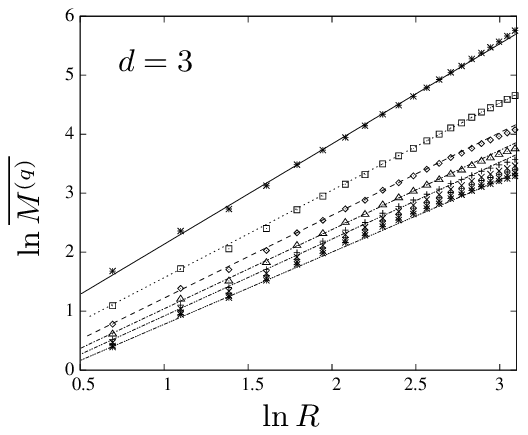}\\
\hspace*{0.1cm}
\includegraphics[width=4.3cm,height=3.7cm]{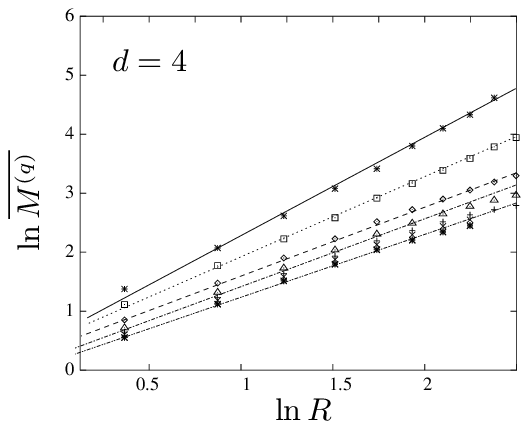}
\end{center}
\caption{\label{moments} Averaged moments ${\overline{M^{(q)}}}$ as function of $R$ in double logarithmic scale in $d=2, 3, 4$. 
In each case, $q=0,$ $1,2,3,4,\ldots$ going from above. Lines are guides to the eyes.}
 \end{figure}

\begin{figure}[htbp]
 \begin{center}
\includegraphics[width=4.1cm,height=3.7cm]{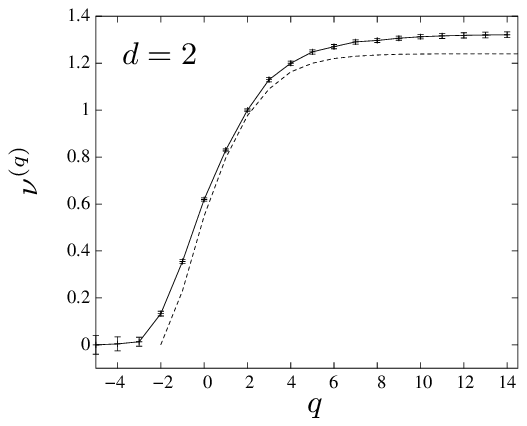}
\hspace*{0.2cm} 
\includegraphics[width=4cm,height=3.7cm]{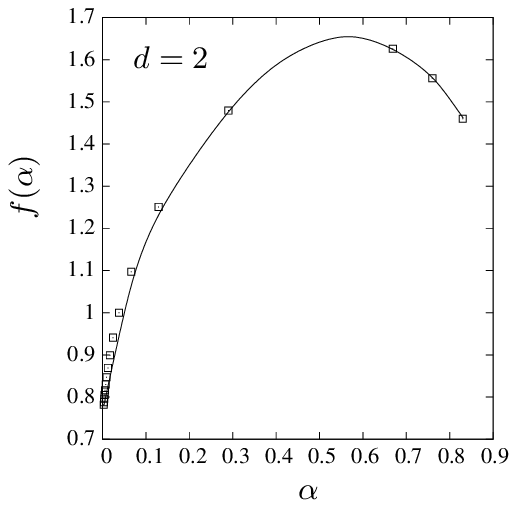}
\includegraphics[width=4.1cm,height=3.7cm]{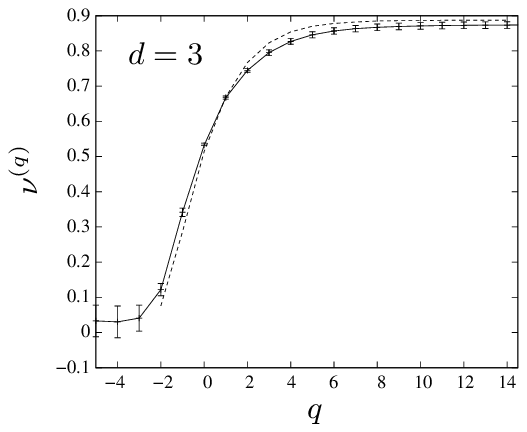}
\hspace*{0.2cm} 
\includegraphics[width=4cm,height=3.7cm]{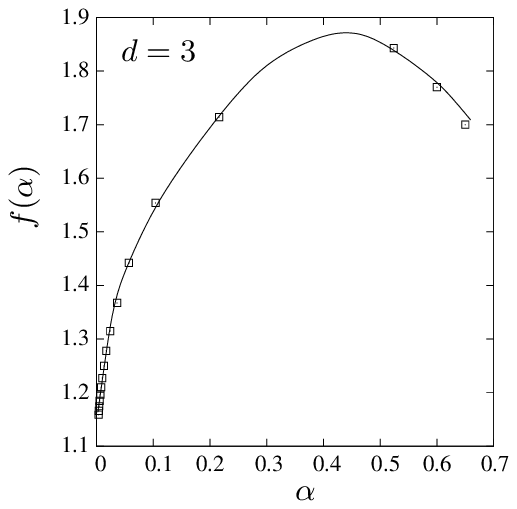}
\includegraphics[width=4.1cm,height=3.7cm]{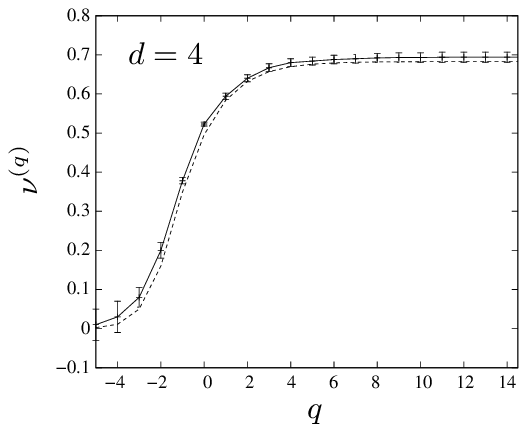}
\hspace*{0.2cm} 
\includegraphics[width=4cm,height=3.8cm]{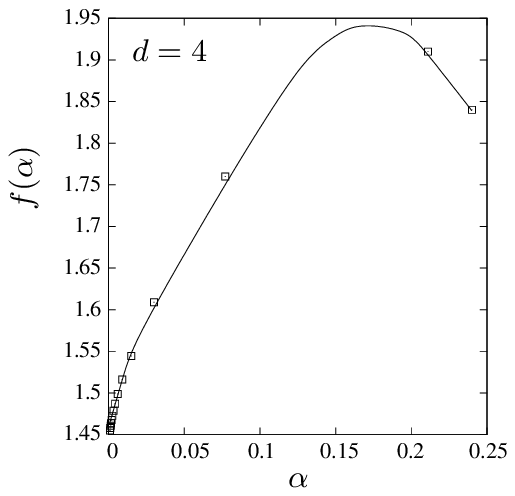}
\end{center}
\caption{\label{nuq} Left: spectrum of multifractal exponents $\nu^{(q)}$ as function of $q$ in $d=2, 3, 4$, dotted lines present 
$[1]/[2]$ Pad\'e approximants to the analytical results 
of Janssen and Stenull (Eq.~(\ref{spectr})). Right: Spectral function $f(\alpha)$ in $d=2, 
3, 4$, the maximum value of $f(\alpha)$ gives the fractal dimension of the underlying backbone of percolation cluster.}
 \end{figure}

The multifractal moments $M^{(q)}$ are defined as follows:
\begin{equation}
M^{(q)}=\sum\limits_{i \in R} {w(i)}^q.
\end{equation}
Averaged over different configurations of the constructed backbones of percolation clusters, they scale as:
\begin{equation} 
{\overline{M^{(q)}}}\sim R^{1/\nu^{(q)}},
\label{momexp}
\end{equation}
with exponents $\nu^{(q)}$  that do not depend on $q$ in  a linear or affine fashion, implying that SAWs on percolation clusters are multifractals.
To estimate the numerical values of  $\nu^{(q)}$ on the basis of data obtained by us (see Fig.~\ref{moments}), 
linear least-square fits are used.
The $\chi^2$ value (sum of squares of normalized deviation from the regression line)  serves as a test of the goodness of fit.

It is clear that at $q=0$ we just count the number of sites of the cluster of linear size $R$, and thus $1/\nu^{(0)}$ corresponds to the fractal 
dimension of the backbone $d_{p_c}^B$. Our results give $d_{p_c}^B(d{=}2)=1.647\pm0.006$, $d_{p_c}^B(d{=}3)=1.865\pm0.006$, $d_{p_c}^B(d{=}4)=1.946\pm0.006$, in 
very good agreement with Table~1.  
At $q=1$, we restore the value of the exponent $\nu_{p_c}$, governing the scaling law of the end-to-end distance for SAWs on the backbone of percolation 
clusters. We obtain $\nu^{(1)}(d{=}2)=0.779\pm0.006$, $\nu^{(1)}(d{=}3)=0.669\pm0.006$, $\nu^{(1)}(d{=}4)=0.591\pm0.006$, in perfect agreement with our recent estimates 
\cite{Blavatska08}
based on the scaling of the end-to-end distance with the number of SAW steps. 
At $q\to \infty$, the so-called ``red sites" of the backbone are mainly taken into account -- the singly connected sites, such that cutting off one of them will 
produce a disconnection of the cluster.  These sites are most often visited by SAWs, and thus have the maximum weights.
It was proven \cite{Coniglio81}, that the number of ``red sites" scales with linear distance $R$ of a cluster as: $
N_{{\rm red}}(R)\sim R^{1/\nu_p}$.
Indeed, as it follows from our data, at large $q$ the value of the exponent $\nu^{(q)}$ tends to $\nu_p$, the percolation correlation length exponent, cf. Table~\ref{dim}.  So, in two limiting cases ($q\to0, q\to \infty$) we restore the corresponding multifractal exponents of the voltage distribution on the backbone of percolation clusters \cite{Arcangelis85}.   
Note that in the latter problem, the exponent governing the scaling of the moment with $q=1$ gives the resistance exponent: e.g., $\nu_R^{(1)}(d{=}2)\simeq0.97$ \cite{Arcangelis85}.
 
Due to the long tail of the distribution $n(w)$, the precision of our estimates decreases with increasing $q$. This problem  turns out to be especially 
crucial when exploring the moments with negative powers $q$: the sites with small probabilities to be visited, which are determinant in negative moments, are very difficult to probe.  

Our estimates of the exponents $\nu^{(q)}$ for different $q$ are presented in the left panel of Fig.~\ref{nuq}. These values appear to be in an 
astonishingly perfect correspondence 
with analytical estimates down to $d=2$ dimensions, 
derived by  applying Pad\'e  approximation to the $\varepsilon=6-d$-expansion (\ref{spectr}), presenting the given series as 
ratio $[m]/[n]$ of two polynomials of degree $m$ and $n$ in $\varepsilon$.
We used the $[1]/[2]$ approximant, because it appears to be most reliable in restoring the known 
estimates in limiting cases ($q=0$, $q\to\infty$). A direct use of the  expression (\ref{spectr}) gives worse results, especially for low dimensions $d$ where the 
expansion parameter $\varepsilon=6-d$ is large.

It is well known that the set of exponents  governing scaling of multifractal moments of the type  (\ref{momexp}) is related to the spectrum  
of singularities $f(\alpha)$ of fractal measure \cite{Halsey86}, called also the  spectral function. 
The physical meaning of $f(\alpha)$ in our problem is that the number $N_R(\alpha)$ of sites $i$, where 
the weight $w(i)$ scales as $R^{-\alpha}$, behaves as: 
\begin{equation}N_R(\alpha) \sim R^{f(\alpha)}.
\end{equation} 
The singularity spectrum  $f(\alpha)$ is given by the Legendre transform:
\begin{equation}
f(\alpha)=q\alpha -\tau(q),\,\,\,\,\alpha(q)=\frac{{\rm d} \tau(q)}{{\rm d} q},
\end{equation}
with $\tau(q)=1/\nu^{(q)}$. Spectral functions $f(\alpha)$, obtained on the basis of our results, are given in the right panel of Fig.~\ref{nuq}. The general 
properties of $f(\alpha)$ are as follows: it is positive on an interval $[\alpha_{{\rm min}},\alpha_{{\rm max}}]$, where $\alpha_{{\rm min}}=\lim\limits_{q\to+\infty}\tau(q)/(q-1)$, $\alpha_{{\rm max}}=\lim\limits_{q\to-\infty}\tau(q)/(q-1)$. The maximum value of the spectral function gives the fractal dimension of the underlying 
structure, which in our case corresponds to the dimension of the backbone of percolation clusters.    

To conclude, we have shown numerically that SAWs residing on the backbone of percolation clusters give rise to a whole spectrum of singularities, 
thus revealing multifractal 
properties. To completely describe peculiarities of the model, the multifractal scaling should be taken into account. We have found estimates for 
the exponents, 
governing different moments of the weight distribution, which scale independently, in surprisingly good coincidence with two-loop $\varepsilon$-expansions. 
The behaviour of the spectral function, describing the frequency of observation of a set of singularities on the underlying backbone 
of percolation clusters, is analyzed as well. 

{\bf Acknowledgement:} V.B. is grateful for support through the Alexander von Humboldt foundation. We thank B. Waclaw for useful discussions.


\begin{thebibliography}{50}

 \bibitem{Stanley88}
For review, see e.g. H.E. Stanley and P. Meakin, Nature {\bf 335}, 405 (1988).


\bibitem{Mandelbrot74}
 B.B. Mandelbrot, J. Fluid Mech. {\bf 62}, 33 (1974).


\bibitem {Witten81}
T.A. Witten and L.M. Sander, Phys. Rev. Lett. {\bf 47}, 1400 (1981).

\bibitem{Arcangelis85}
 L. de Arcangelis, S. Redner, and A. Coniglio, Phys. Rev. B {\bf 31}, 4725 (1985); 
 Phys. Rev. B {\bf 34}, 4656 (1986); 
   R. Rammal, C. Tannous,  and A.M.S. Tremblay, Phys. Rev. A {\bf 31}, 2662 (1985);
 R. Blumenfeld and A. Aharony, J. Phys. A {\bf 18}, L443 (1985).

\bibitem{Halsey86}
 T.C. Halsey, P. Meakin, and I. Procaccia, Phys. Rev. Lett. {\bf 56},  854 (1986);  T.C. Halsey, M.H. Jensen, L.P. Kadanoff,  I. Procaccia, and  B.I. Shraiman,   Phys. Rev. A
 {\bf 33}, 1141 (1986).

\bibitem{Jensen02}
M.H. Jensen et al., Phys. Rev. E {\bf 65}, 046109 (2002);  M.H. Jensen, J. Mathiesen, and I. Procaccia, Phys. Rev. E {\bf 67}, 042402 (2003).



\bibitem{Hentschel83}
 H.G. Hentschel and I. Procaccia, Physica D {\bf 8}, 435 (1983);
R. Benzi et al., J. Phys. A {\bf 17}, 3521 (1984); A. Brandenburg et al., Phys. Rev. A {\bf 46}, 4819 (1992); D. Queiros-Conde, Phys. Rev. E {\bf 64},
 015301 (2001);  L. Biferale et al., Phys. Rev. Lett. {\bf 93}, 064502 (2004).
  
\bibitem{Ivanov99}
P. Ch. Ivanov et al., Nature {\bf 399}, 461 (1999).

\bibitem{Schreiber91}
M. Schreiber and H. Grussbach, Phys. Rev. Lett. {\bf 67}, 607 (1991); H. Grussbach and M. Schreiber, Phys. Rev. B {\bf 51}, 663 (1995);
A. Mildenberger, F. Evers, and A.D. Mirlin, Phys. Rev. B {\bf 66}, 033109 (2002); A. D. Mirlin et al., Phys. Rev. Lett. {\bf 97}, 046803 (2006).

  
\bibitem{Dhar78}
D. Dhar, J. Math. Phys. {\bf 19}, 5 (1978); J. Physique {\bf 49}, 397 (1988);
  I. \^Zivi\`c, S. Milo\^sevi\`c, and B. Djordjevi\`c, J. Phys. A {\bf 38}, 555 (2005); A. Ordemann, M. Porto, and H.E. Roman, Phys. Rev. E {\bf 65}, 021107 (2002); 
F.D. Reis and R. Riera, J. Phys. A {\bf 28}, 1257 (1995);
A. Ordemann, M. Porto, and H.E. Roman, J. Phys. A {\bf 35}, 8029 (2002).    


\bibitem{Kremer81}
K.~Kremer, Z. Phys. B {\bf 45}, 149 (1981);
S.B. Lee and H. Nakanishi, Phys. Rev. Lett. {\bf 61}, 2022 (1988); S.B. Lee, H. Nakanishi, and Y. Kim, Phys. Rev. B {\bf 39}, 9561 (1989); 
K.Y. Woo and S.B. Lee, Phys. Rev. A {\bf 44}, 999 (1991); S.B. Lee, J. Korean Phys. Soc. {\bf 29}, 1
(1996); H. Nakanishi and S.B. Lee, J. Phys. A {\bf 24}, 1355 (1991).

\bibitem{Meir89}
Y. Meir and A.B. Harris,   Phys. Rev. Lett. {\bf 63},  2819  (1989).

\bibitem{Grassberger93}
P.~Grassberger, J. Phys. A {\bf 26}, 1023 (1993).

\bibitem{Rintoul94}
M.D. ~Rintoul, J.~Moon, and H.~Nakanishi, Phys. Rev. E {\bf 49},
2790 (1994).

\bibitem{Ordemannsaw}
A.~Ordemann et al., Phys.
Rev. E {\bf 61}, 6858 (2000).

\bibitem{Blavatska08}
V. Blavatska and W. Janke, Europhys. Lett. {\bf 82}, 66006 (2008).

\bibitem{Kim83}
Y. Kim, J. Phys. C {\bf 16}, 1345 (1983); J. Phys. A {\bf 20}, 1293 (1987).

\bibitem{Barat91}
K. Barat, S.N. Karmakar, and B.K. Chakrabarti, J. Phys. A {\bf 24}, 851 (1991). 

\bibitem{Blavatska04}
C. von Ferber et al.,  Phys.
Rev. E  {\bf 70},  035104(R) (2004).

\bibitem{Janssen07}
H.-K. Janssen  and O. Stenull, Phys. Rev. E {\bf 75}, 020801(R)(2007).


\bibitem{Havlin87}
S. Havlin  and D. Ben Abraham, Adv. Phys. {\bf 36},  {155} (1987).

\bibitem{Grassberger86}
P. Grassberger,  J. Phys. A {\bf 19},  {1681} (1986).


 \bibitem{Moukarzel98}
C. Moukarzel, Int. Journ.  Mod. Phys. C {\bf 8},   887 (1998).

\bibitem{Nijs79}
 M.P.M. den Nijs, J. Phys. A {\bf 12}, 1857 (1979);
 B. Nienhuis, J. Phys. A {\bf 15}, 199 (1982).
 
%
 \bibitem{Strenski91}
 P.N. Strenski, R.M. Bradley, and J.M. Debierre, Phys. Rev. Lett. {\bf 66}, 1330 (1991).
 
 
 \bibitem{Janssen00}
 H.K. Janssen and O. Stenull, Phys. Rev. E {\bf 61}, 4821 (2000).
 
 \bibitem{Amit76}
 D.J. Amit, J. Phys. A {\bf 9}, 1441 (1976).
 
\bibitem{Porto97}
M. Porto et al., Phys. Rev. E {\bf 56}, 1667 (1997).


\bibitem{Havlin84}
S. Havlin et al., J. Phys. A {\bf 17}, L957 (1984).
 
 
 \bibitem{Rosenbluth55}
M.N. Rosenbluth  and A.W. Rosenbluth, J. Chem. Phys. {\bf 23}, {356} (1955);
P. Grassberger, Phys. Rev. E {\bf 56}, 3682 (1997).


 \bibitem{Coniglio81}
 A. Coniglio, Phys. Rev. Lett. {\bf 46}, 250 (1981); J. Phys. A {\bf 15}, 3829 (1982).
 
 
  
\end{thebibliography}
\end{document}